\documentclass[prd,preprint,eqsecnum,nofootinbib,amsmath,amssymb,preprintnumbers,tightenlines,longbibliography]{revtex4-1}
\usepackage[colorlinks,linkcolor=blue,citecolor=blue]{hyperref}
\usepackage{amsmath,graphicx}
\usepackage{color}
\usepackage[dvipsnames]{xcolor}
\definecolor{mygray}{gray}{0.6}

\def\t{\tilde}

\def\E{{\mathcal E}}

\def\P{{\mathcal P}}

\def\mn{{\mu\nu}}

\def\Eq#1{Eq.~(\ref{#1})}
\def\Eqs#1{Eqs.~(\ref{#1})}

\def\Fig#1{Fig.~\ref{#1}}
\def\Sect#1{Section~\ref{#1}}
\def\Ref#1{Ref.~\cite{#1}}

\def\bra{\langle}
\def\ket{\rangle}

\def\ie{\emph{i.e.}}
\newcommand{\qbra}[1]{{\langle #1 \vert}}
\newcommand{\qket}[1]{{\vert #1\rangle}}
\newcommand \beq{\begin{eqnarray}}
\newcommand \eeq{\end{eqnarray}}
\newcommand \be{\begin{equation}}
\newcommand \ee{\end{equation}}

\begin{document}

\title{Hydrodynamic attractor in the non-conformal Bjorken flow}

\author{Zenan Chen}
\author{Li Yan} 
\email{cliyan@fudan.edu.cn}
\affiliation{
Key Laboratory of Nuclear Physics and Ion-Beam Application (MOE) \& Institute of Modern Physics at
Fudan University,\\
220 Handan Road, 200433, Yangpu District, Shanghai, China
}
\date{\today}

\begin{abstract}

Non-conformal attractor behavior is studied by solving non-conformal second order viscous hydrodynamics with respect to boost-invariant plasmas. Numerical solutions of the relative decay rate of the enthalpy density, the inverse shear and bulk Reynolds numbers, all exhibit universal patterns towards an ideal fluid description at late stages, demonstrating the existence of non-conformal hydrodynamic attractor. However, as a generic feature, the non-conformal hydrodynamic attractor emerges only at late times, which differs drastically from the behavior observed in conformal systems. The absence of the early-time attractor in non-conformal fluids is a consequence of the early-time instable mode, which arises from a positive-valued eigen-mode associated with the free-streaming dynamics. To support practical simulations of hydrodynamics in high-energy heavy-ion collisions, early-time attractor can be restored in a mixture of the shear and the bulk viscous corrections, in which early-time instable modes cancel approximately. Early-time non-conformal attractor is also approachable, if the quadratic coupling between the bulk pressure and the expansion rate is enhanced by at least a factor of two through the transport coefficient $\delta_{\Pi\Pi}$.

\end{abstract}
\maketitle

\section{introduction}


The fluid nature of quark-gluon plasmas (QGP)  is one of the most significant discoveries achieved in heavy-ion experiments at Relativistic Heavy-Ion Collider (RHIC) and the Large Hadron Collider (LHC)~\cite{Shuryak:2003xe}. It has been extensively studied through the measurements of collective flow phenomena among the created particles. These observed flow phenomena from various colliding systems are consistent with 
those obtained by hydrodynamical modelings of QGP expansion in the corresponding systems~\cite{Heinz:2013th,Gale:2013da}. 

Despite the success of hydrodynamic modelings, it is still a theoretical question how hydrodynamics starts to dominate the QGP expansion shortly 
after the initial collisions between nucleus and nucleus, proton and nucleus or even proton and proton. 
As an effective theory characterizing the long-wavelength and low-frequency excitations, hydrodynamics is often formulated in terms of expansion in gradients. Accordingly, application of hydrodynamics requires that the spatial and temporal gradients to be small, so that truncation of the gradient expansion at finite orders suffices to give reliable descriptions.  However, in the very early stages of heavy-ion collisions, 
the approximately boost-invariant geometry of the QGP medium implies large spatial gradients, 
which apparently violates the applicability condition of hydrodynamics. This discrepancy between the theory of hydrodynamics and phenomenology was recently solved in terms of the so-called attractor solution for \emph{conformal systems}~\cite{Heller:2015dha,Romatschke:2017vte,Romatschke:2017ejr}. In such solutions, initial condition independent evolution emerges as an universal pattern before the dominance of interaction over longitudinal expansion, at which the system is still far from local equilibrium, and approaches ideal hydrodynamics at late times.

Conformal attactors have been obtained in kinetic theory~\cite{Heller:2016rtz,Kurkela:2019set,Brewer:2019oha,Blaizot:2017ucy,Almaalol:2020rnu,Strickland:2018ayk,Kamata:2020mka,Behtash:2020vqk,Behtash:2018moe} and in the solution of second order dissipative fluid dynamics~\cite{Kurkela:2019set,Basar:2015ava,Jaiswal:2019cju,Jaiswal:2019hen,Chattopadhyay:2019jqj,Strickland:2017kux,Heller:2020anv}. From both formulations, the resulted attractors share quantitative similarity that captures the system evolution towards hydrodynamization, albeit some quantitative difference do exist in the early-time or free-streaming extreme~\cite{Blaizot:2021cdv}.  
Moreover, the early-time power-law decay and the late-time exponential decay of initial condition dependent evolutions are found generic in both formulations. In particular, the early-time power-law decay is essential to the early-time conformal attractor, 
which allows one to initialize hydrodynamics in practice starting from the very early stages, 
while the subsequent out-of-equilibrium evolution 
is deterministic and describable by the hydrodynamic attractor. Therefore, 
the early-time attractor behavior is an key factor for the applications of hydrodynamics with early-time initialization to heavy-ion collisions, especially for the medium created in small colliding systems.

With respect to conformal boost-invariant plasmas, the attractor behavior of the system evolution can be well understood in different respects~\cite{Shen:2020mgh}. First, the system evolution towards local equilibrium can be recognized as a transition from the free-streaming expansion to an ideal fluid. The free-streaming expansion dominates in the free-streaming extreme, which contains a couple of fixed points~\cite{Blaizot:2017ucy,Blaizot:2019scw}. Attractor solution is then initialized at the stable free-streaming fixed point, evolving towards the ideal fluid pseudo-fixed point which is trivially known. Secondly, when hydrodynamic equations of motion are written as a coupled mode problem, approximately, attractors can be seen as the evolution of slow modes, in the free-streaming regime and in the hydrodynamic regime respectively~\cite{Brewer:2019oha}. Thirdly, hydrodynamic gradient expansion in the boost-invariant plasmas is asymptotic~\cite{Heller:2013fn,Heller:2015dha,Basar:2015ava,Blaizot:2017lht,Heller:2020uuy}. 
Borel summation can be applied to the asymptotic hydrodynamic series which results in a trans-series. Resummation of the trans-series allows one to identify the attractor solution~\cite{Heller:2015dha,Basar:2015ava,Blaizot:2020gql}.  Interestingly, from the Borel summation technique, initial condition dependent solutions are found to be transient against the conformal attractor, which exhibit power-law decay at early times and exponential decay at late times~\cite{Kurkela:2019set}.

Although the conformal attractor has been extensively analyzed, conformal symmetry in realistic QGP media in high-energy heavy-ion collisions is actually broken, for instance, by the small masses of up- and down-quarks. Therefore, in realistic QGP, in addition to the shear viscous corrections, bulk viscous effects are expected which may alter the established picture of hydrodynamic attractor, and accordingly the applicability of hydrodynamics to out-of-equilibrium plasmas. In fact, some recent studies have shown that 
the early-time attractor is not available in non-conformal systems~\cite{Chattopadhyay:2021ive,Jaiswal:2021uvv}, which hinders the application of non-conformal fluid dynamics to the realistic QGP. 
The purpose of the current study is to investigate the applicability of non-conformal dissipative fluid dynamics to realistic QGP media out of equilibrium, based on the properties of non-conformal attractor.  It is also our 
interest to examine how the breaking of conformal sysmmetry modifies the behavior of attractor solution. 

The paper is organized as follows. With the formulation of non-conformal viscous hydrodynamics given with respect to a boost-invariant plasma in \Sect{sec:2}, non-conformal attractors are solved and studied in \Sect{sec:3}. Although the non-conformal attractor can be analogously analyzed via the fixed points or slow-mode evolution, we shall concentrate in this work on the non-conformal gradient expansion, as in \Sect{sec:gexp}. Perturbations around the gradient expansion are discussed in \Sect{sec:modes}, where we also introduce the early-time stable and instable modes. In \Sect{sec:4}, we propose two possible strategies that help restore the early-time non-conformal attractor. Summary and discussions are given in \Sect{sec:5}.

\section{Non-conformal fluid dynamics and Bjorken flow}
\label{sec:2}

Hydrodynamics is an effective theory for low-frequency and long-wavelength excitations. Evolution of these excitations can be  captured by solving conservation laws with respect to a set of hydrodynamical variables, including the energy density $\E$, the pressure $\P$ and flow four velocity $u^\mu$. The conservation of energy and momentum, for instance,
\be
\label{eq:eom0}
\partial_\mu T^\mn=0\,,
\ee
is solved through the energy-momentum tensor $T^\mn$ and a constitutive relation that relates $T^\mn$ to the hydrodynamical variables and their gradients.

The standard formulation of the hydrodynamic constitutive relation applies to systems close to local thermal equilibrium, where the off-equilibrium effects are dissipations. In hydrodynamics, these dissipations are treated in terms of an expansion of spatial and temporal gradients of the hydrodynamical variables order by order. 
For the energy-momentum tensor, in addition to the ideal fluid description, there are contributions from the shear stress tensor $\pi^\mn$ and the bulk viscous pressure\footnote{
A mostly negative matrix signature is implied in the formula, $g^{\mu\nu}=(+,-,-,-)$. Accordingly, the projection tensor is $
\Delta^{\mu\nu} =  g^{\mu\nu}-u^\mu u^\nu $, with which the spatial gradient $\nabla^\mu = \Delta^{\mu\nu} \partial_\nu$ reduces to $\partial^i$ ($i=x,y,z$) in the local rest frame.
}, 
\be
T^{\mu\nu} = 
\E u^\mu u^\nu - (\P+\Pi)\Delta^\mn + \pi^\mn\,.
\ee
In the leading order, the structure of these tensors are given by the Navier-Stokes (NS) hydrodynamics,
\be
\label{eq:NShydro}
\pi^{\mu\nu} = \eta ^\bra\nabla^\mu {u^\nu}^\ket + O(\nabla^2)\,,\qquad
\Pi = -\zeta \nabla\cdot u + O(\nabla^2)\,,
\ee
where the shear viscosity $\eta$ and the bulk viscosity $\zeta$ are the corresponding leading order transport coefficients. In \Eq{eq:NShydro} and in what follows, the angular brackets around tensor indices are used to indicate tensors being symmetric, transverse to flow four velocity and traceless. Hydrodynamic gradient expansion must be truncated to finite orders for practical applications. In particular, when truncating to the second order in spatial gradients, a casual formulation can be realized which relaxes to the NS relations in \Eq{eq:NShydro}. This is the well-known Mueller-Isreal-Stewart (MIS) hydrodynamics, which is the essential ingredient of the 
modeling of QGP evolution in relativistic heavy-ion collisions~\cite{Gale:2013da}. 

One often used non-conformal generalization of the MIS hydrodynamics has the following equations of motion for $\pi^{\mn}$ and $\Pi$~\cite{Denicol:2014vaa,Ryu:2017qzn},
\begin{subequations}
\label{eq:dnmr}
\begin{align}
\tau_\pi \,^\bra {u^\alpha\partial_\alpha\pi^\mn}^\ket+\pi^\mn =&\;
2\eta \sigma^\mn -\delta_{\pi\pi} \pi^\mn \nabla\cdot u 
- \tau_{\pi\pi} \pi^{\bra\mu}_{\;\;\;\;\alpha} \sigma^{\nu\ket\alpha} + \lambda_{\pi\Pi} \Pi \sigma^\mn\,, \\
\tau_\Pi u^\alpha \partial_\alpha \Pi + \Pi =& - \zeta \nabla\cdot u - \delta_{\Pi\Pi} \Pi \nabla\cdot u + \lambda_{\Pi\pi} \pi^{\mn}\sigma_{\mn}\,,
\end{align}
\end{subequations}
with $\sigma^{\mn}= 2^\bra\nabla^\mu {u^\nu}^\ket$. In \Eqs{eq:dnmr}, like $\eta$ and $\zeta$, all second order transport coefficients are input parameters that should be determined by the corresponding underlying microscopic dynamics. For instance, in a kinetic theory framework~\cite{York:2008rr}, the shear relaxation time is found to be proportional to the ratio between the shear viscosity and the enthalpy density, $\tau_\pi\propto\eta/(\E+\P)$. Except $\tau_\pi$ and $\tau_\Pi$, all other second transport coefficients in \Eqs{eq:dnmr} arise from quadratic couplings of viscous corrections. For instance, $\lambda_{\pi\Pi}$ and $\lambda_{\Pi\pi}$ are associated with the effect of shear-bulk coupling~\cite{Denicol:2014mca}. The coefficient $\delta_{\Pi\Pi}$ controls the coupling between the bulk viscous pressure and the local expansion rate $\nabla\cdot u$. These transport coefficients of non-linear viscous corrections are more sensitive to the out-of-equilibrium system evolution where gradients are large.

At the early stages in high-energy heavy-ion collisions, the created QGP medium is approximately boost invariant along the longitudinal direction, namely, the direction along the beam axis ($z$-axis), while expansion in the transverse plane ($\vec x_\perp$-plane) is negligible. Accordingly, system evolution can be simplified as the so-called Bjorken flow. In particular, in the Milne coordinates of $(\tau,\xi)$, 
\be
\tau = \sqrt{t^z-z^2}\,,\qquad
\xi = \tanh^{-1} \left(\frac{z}{t}\right)\,,
\ee
dependence on the space-time rapidity $\xi$ is suppressed with respect to the Bjorken symmetry, and the system evolution relies only the proper time $\tau$. As a consequence of the Bjorken flow, the flow four velocity is known as well, $u^\mu=(1,\vec 0)$, which further leads to $\nabla^\xi u_\xi \sim \nabla\cdot u \sim 1/\tau$. Given all these simplifications, hydrodynamic equations of motion, \Eq{eq:eom0} and \Eqs{eq:dnmr} reduce to coupled ordinary differential equations (ODE) with respect to the proper time $\tau$,
\begin{subequations}
\label{eq:eomBjork}
\begin{align}
\frac{d \E}{d\tau} & = - \frac{1}{\tau}(\E + \P + \Pi - \pi) \,,\\
\tau_\pi \frac{d\pi}{d\tau} + \pi & = \frac{4}{3} \frac{\eta}{\tau} - \left(\frac{1}{3}\tau_{\pi\pi}+ \delta_{\pi\pi}\right) \frac{\pi}{\tau} + \frac{2}{3} \lambda_{\pi\Pi} \frac{\Pi}{\tau}\,,\\
\tau_\Pi \frac{d \Pi}{d\tau} + \Pi & = - \frac{\zeta}{\tau} - \delta_{\Pi\Pi} \frac{\Pi}{\tau} + \lambda_{\Pi\pi} \frac{\pi}{\tau}\,,
\end{align}
\end{subequations}
where $\pi=\pi^\xi_{\;\xi}$ is the only non-zero component of the shear stress tensor.

To solve \Eqs{eq:eomBjork}, an equation of state (EoS) must be provided to relate the hydrodynamical variables. 
For instance, energy density and pressure are related via the velocity of sound, $c_s^2=\partial\P/\partial \E$. In general, $c_s^2$ depends on energy density which makes the relation nonlinear. When conformal symmetry is slightly broken by a tiny but finite quark mass, especially comparing to the characteristic energy scale $Q$ in the sytem, so that $z_*\equiv m/Q\ll 1$, as an approximation one may consider instead a linear equation of state~\cite{Jaiswal}, 
\be
\P=c_s^{2} \E + O(z_*)\,,
\ee
where $c_s^2$ is a constant. This is indeed the case in the early stages of high-energy heavy-ion collisions, where the energy scale of QGP is the saturation scale $Q_s\sim O(1)$ GeV and $z_*=m/Q_s\sim 10^{-3}$. 
With repect to extremely high energy scales, the breaking of conformal symmetry in QGP also results in a small deviation of the speed of sound from 
its confomal limit, from which another small quantity can be introduced, $0<\epsilon\equiv 1/3-c_s^2\ll 1$. At sufficiently high energy densities, lattice QCD calculations show that $\epsilon\sim 10^{-2}$ (cf. \Ref{Ding:2015ona}). Hereafter, we shall always treat $z_*$ and $\epsilon$ as the two small  parameters that break the conformal symmtry.

\section{Pre-equiliurm evolution of non-conformal QGP}
\label{sec:3}

\subsection{Coupled mode formulation}

With the help of the linear relation between energy density and pressure, \Eqs{eq:eomBjork} can be recast into a coupled mode problem for $(\E+\P, \pi, \Pi)$,
\begin{subequations}
\label{eq:eom1}
\begin{align}
\label{eq:eom1a}
\tau\frac{d(\E+\P)}{d\tau} &= - (1+c_s^2)(\E + \P + \Pi - \pi) \,,\\
\tau\frac{d\pi}{d\tau}  & 
=\alpha_\pi(\E+\P)-\gamma_1 \pi + \gamma_2 \Pi - w_\pi \pi\,,\\
 \tau\frac{d \Pi}{d\tau} & 
=-\alpha_\Pi(\E+\P) + \gamma_3 \pi-\gamma_4 \Pi - w_\Pi \Pi\,,
\end{align}
\end{subequations}
where 
\be
w_\pi = \frac{\tau}{\tau_\pi}\,,\qquad
w_\Pi = \frac{\tau}{\tau_\Pi}
\ee
are the inverse Knudsen numbers in the shear and the bulk channels, respectively. 
\Eqs{eq:eom1} are accurate up to the order $O(z_*^{0})$. Especially, as we have introduced the dimensionless first order transport coefficients~\cite{Ryu:2017qzn},
\begin{align}\label{eq:trans1}
\alpha_\pi = \frac{4\eta}{3\tau_\pi(\E+\P)}=\frac{4}{15}+O(z_*^{2})\,,\qquad
\alpha_\Pi = \frac{\zeta}{\tau_\Pi(\E+\P)}=14.55\epsilon^2+O(z_*^{5})\,,
\end{align}
and second order transport coefficients,
\begin{align}\label{eq:trans2}
\gamma_1 &= \frac{1}{\tau_\pi}\left(\frac{1}{3}\tau_{\pi\pi}+ \delta_{\pi\pi}\right)=\frac{38}{21}+O(z_*^{2})\,,
&&\gamma_2 = \frac{2}{3} \frac{\lambda_{\pi\Pi}}{\tau_\pi}=\frac{4}{5}+O(z_*^{2}\ln z_*)\,,\nonumber\\
\gamma_3 &= \frac{\lambda_{\Pi\pi}}{\tau_\Pi}=\frac{8}{5}\epsilon+O(z_*^{4})\,,
&&\gamma_4 = \frac{\delta_{\Pi\Pi}}{\tau_\Pi}=\frac{2}{3}+O(z_*^{2}\ln z_*)\,,
\end{align}
and consider only the $O(z_*^0)$ constant values of these coefficients for analysis. 
In the conformal limit, namely, when $z_*\to 0$ and $\epsilon\to0$ simultaneously, one notices that $\Pi\to0$, $\alpha_\Pi\to0$ and $\gamma_3\to0$, with which \Eqs{eq:eom1} reduce to the conformal equations of motion for $\E+\P$ and $\pi$. 

Solving \Eqs{eq:eom1} allows one to investigate the pre-equilibrium evolution of a non-conformal QGP, and to study the onset of non-conformal hydrodynamics. Before proceed, several comments are in order. 

First, as the hydrodynamic equations of motion, \Eqs{eq:eom1} are expected valid only in the hydrodynamic regime, \ie, under the condition of small Knudsen numbers so that $w_\pi\sim w_\pi\gg1$. This is the applicability condition of the standard hydrodynamic formulation based on truncated gradients. However, extension of second order viscous hydrodynamics to the far-away-from-equilibrium regime, where $w_\pi\sim w_\Pi \ll1$, as has been acknowledged from the analysis of conformal fluids (cf. \Ref{Basar:2015ava}), is allowed to give descriptions qualitatively correct which contains attractor behavior. In the conformal fluids, the quantitative discrepancy in the attractor solution appears as a mismatch of the free-streaming fixed points, which can be fixed by a renormalized transport coefficient~\cite{Blaizot:2021cdv}. In a similar strategy, we shall solve the non-conformal hydrodynamic equations of motion, in regimes where $w_\pi\sim w_\Pi\ll1$.

Secondly, to solve the coupled equations for realistic non-conformal QGP, one also needs to parameterize the time dependence of relaxation time in both the shear and the bulk channels. We shall consider two constants $\Delta_\pi$ and $\Delta_\Pi$ for $w_\pi$ and $w_\Pi$, respectively, as,
\be
\label{eq:relax}
\begin{cases}
\tau_\pi \propto \tau^{1-\Delta_\pi}\\
\tau_\Pi \propto \tau^{1-\Delta_\Pi}
\end{cases}
\quad
\mbox{and}
\qquad
\begin{cases}
w_\pi \propto \tau^{\Delta_\pi}\\
w_\Pi \propto \tau^{\Delta_\Pi}
\end{cases}
\ee
By varying the values of these constants, one is allowed to mimic a large variety of microscopic interactions among constituents of the medium~\cite{Blaizot:2021cdv,Jaiswal:2019cju}. For instance, for a conformal system, it has been shown that $\tau_\pi\propto 1/T$ and the system evolution can be approximately captured by taking $\Delta_\pi=4/3$. In the current work, for illustrative purpose and for simplicity, we consider a scenario that $\Delta_\pi=\Delta_\Pi=1$, which corresponds to a constant relaxation time in both the shear and the bulk channels, and make the identification $\tau_\pi=\tau_\Pi=\tau_R$. This in practice can be realized in kinetic theory of a constant relaxation time $\tau_R$~\cite{Jaiswal:2021uvv}. Accordingly, the inverse Knudsen numbers in both channels equal, $w_\pi=w_\Pi=w\propto \tau$. Generalization to arbitrary values of $\Delta_\pi$ and $\Delta_\Pi$ is straightforward, which however, does not change qualitatively the results we shall present.

Thirdly, as a coupled mode problem, \Eqs{eq:eom1} can be written with respect to the vector $\qket{L}=(\E+\P, \pi, \Pi)$, as,
\be
\label{eq:coupled}
\tau\partial_\tau |L\ket = -H_0 |L\ket - w H_1 |L\ket
\ee
where the two matrices are
\be
\label{eq:matrix}
H_0
=\begin{pmatrix}
4/3-\epsilon & -(4/3-\epsilon) & 4/3-\epsilon \\
-\alpha_\pi & \gamma_1 & -\gamma_2 \\
\alpha_\Pi & -\gamma_3 & \gamma_4 
\end{pmatrix}\,,\qquad
H_1 = 
\begin{pmatrix}
0 & 0 & 0 \\
0 & 1 & 0 \\
0 & 0 & 1
\end{pmatrix}\,,
\ee
characterizing the effects of free streaming and interaction that dominates at early and late times, respectively. For a system far from local equilibrium, where $w\ll1$, interactions can be neglected and the system expansion is dominated by free streaming. This is the free-streaming limit. In the free streaming limit, conformal fluid has two fixed points. The stable free streaming fixed point results in one special initial condition that generates the attractor solution. In the non-conformal system, however, corresponding to the eigenvalues of $H_0$ in \Eq{eq:matrix}, one extra fixed point appears, associated with bulk viscous corrections. 
More detailed discussions on the non-conformal free-streaming fixed points will be given in \Ref{Jaiswal}.

\subsection{Attractor solutions of non-conformal viscous hydrodynamics}

In conformal fluids, attrator solutions of the system evolution appear in certain combinations of hydrodynamical variables, such as the relative decay rate of energy density, $d\ln\E/d\ln \tau$ and the pressure ratio between the longitudinal and the transverse pressures $\P_L/\P_T$. It has been noticed that, these combinations are all related, in one way or another, to the inverse shear Reynolds number, $\pi/(\E+\P)$. In fact, this only quantity suffices to describe the expansion of a conformal fluid from far from equilibrium to local equilibrium. In particular, it captures the attractor behavior in the evolution. In the non-conformal case, analogously, in order to fully characterize the system evolution and to search for attractor solutions, we consider the following three dimensionless quantities,
\be
\label{eq:three}
g(w) \equiv \frac{\partial\ln(\E+\mathcal{P})}{\partial\ln\tau}\,,\quad
\bar \pi \equiv \frac{\pi}{\E+\P}\,,\quad
\bar \Pi \equiv \frac{\Pi}{\E + \P}\,,
\ee
namely, the relative decay rate of the enthalpy density, the inverse shear and bulk Reynolds numbers. 
Only two of these three quantities are independent, as can be deduced from \Eq{eq:eom1a}, that
\be
\label{eq:gpP}
	g(w)=-(4/3-\epsilon)(1+\bar{\Pi}-\bar{\pi})\,.
\ee 
Comparing to conformal systems, where there is only one independent quantity, the extra degree of freedom comes from the bulk viscous correction. This can be consistently seen, as in the conformal limit, \ie, by taking $\epsilon\to0$ and $\bar \Pi\to 0$, that \Eq{eq:gpP} reduces to a relation that relates the relative decay rate of energy density and the inverse shear Reynolds number, and the pressure anisotropy, as expected, 
\be
g(w) \equiv \frac{d\ln \E}{d\ln \tau} = -\frac{4}{3}(1-\bar \pi) \sim \P_L/\P_T\,.
\ee
In the current work, we shall focus on 
$\bar\pi$ and $\bar\Pi$, while $g(w)$ can be obtained via \Eq{eq:gpP} once $\bar\pi$ and $\bar\Pi$ are given.

Although these quantities can be solved directly from the equation of motion \Eqs{eq:eom1}, it is not difficult to re-derive a coupled first order nonlinear differential equations for $(\bar\pi,\bar \Pi)$, as a function of $w$,
\begin{subequations}
\label{eq:eomc}
\begin{align}
\label{eq:grelation}
	w\frac{\partial\bar{\pi}}{\partial  w}&=\alpha_\pi-\gamma_1\bar{\pi}+\gamma_2\bar{\Pi}-g(w)\bar{\pi}-w\bar{\pi}\,,\\
		w\frac{\partial\bar{\Pi}}{\partial  w}&=-\alpha_\Pi+\gamma_3\bar{\pi}-\gamma_4\bar{\Pi}-g(w)\bar{\Pi}-w\bar{\Pi}\,.
\end{align}
\end{subequations}
\Eqs{eq:eomc} generalize the nonlinear differential equation that has been investigated in the conformal case (cf.~\cite{Basar:2015ava}). In the limiting regimes of $w$, solutions to \Eqs{eq:eomc} can be analytically obtained. For instance, the fixed points of $\bar\pi$ and $\bar\Pi$ in the hydrodynamic limit $w\to\infty$ can be read off as $\bar\pi=\bar\Pi\to 0$, which further gives $g\to -4/3+\epsilon$. In the opposite free-streaming limit, $w\to 0$, the fixed points are solutions to coupled algebraic equations, which coincide with the eigenvalues of the matrix $H_0$ in \Eq{eq:matrix}.


\begin{figure}
\begin{center}
\includegraphics[width=0.315\textwidth] {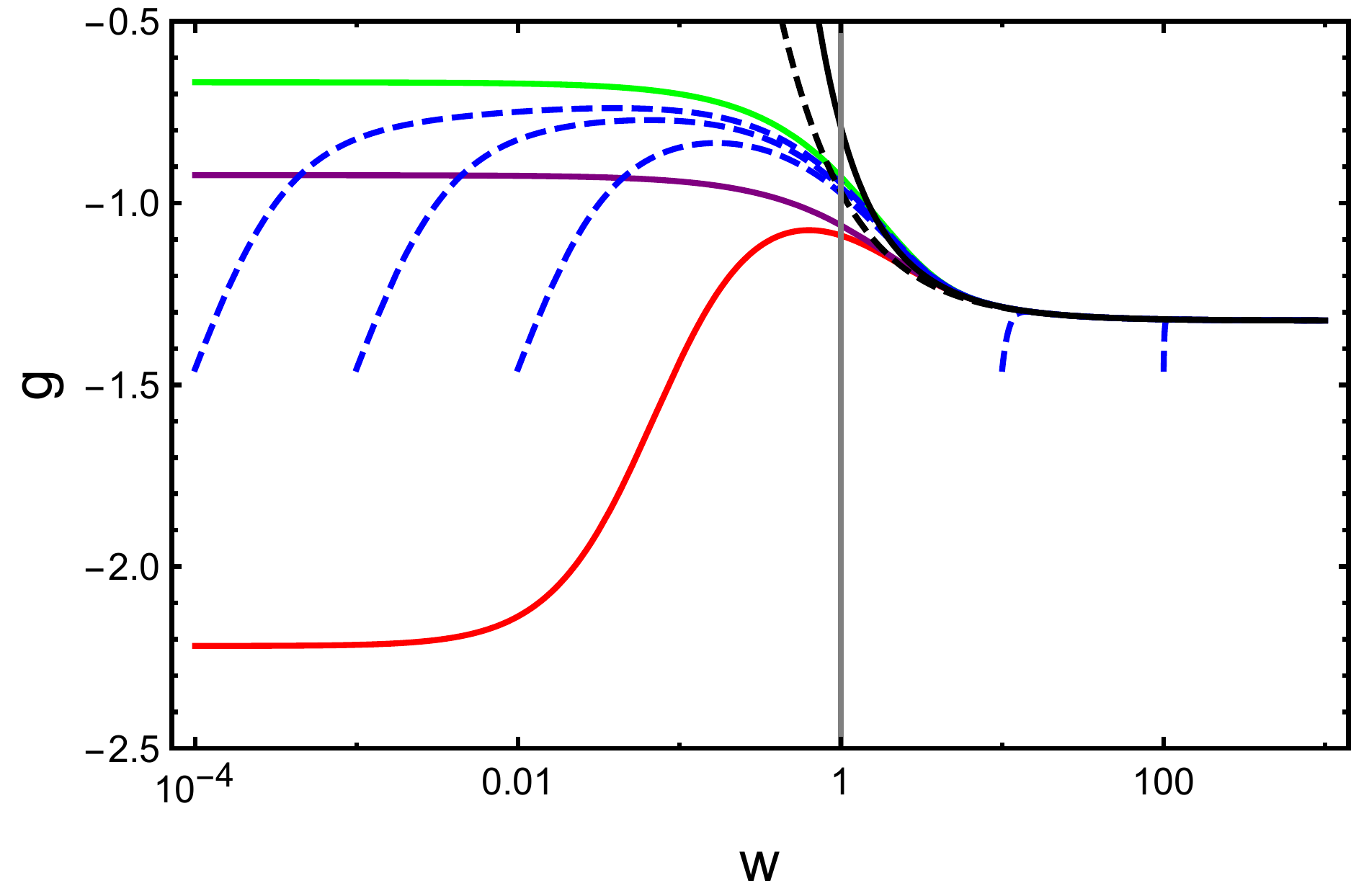}
\includegraphics[width=0.315\textwidth] {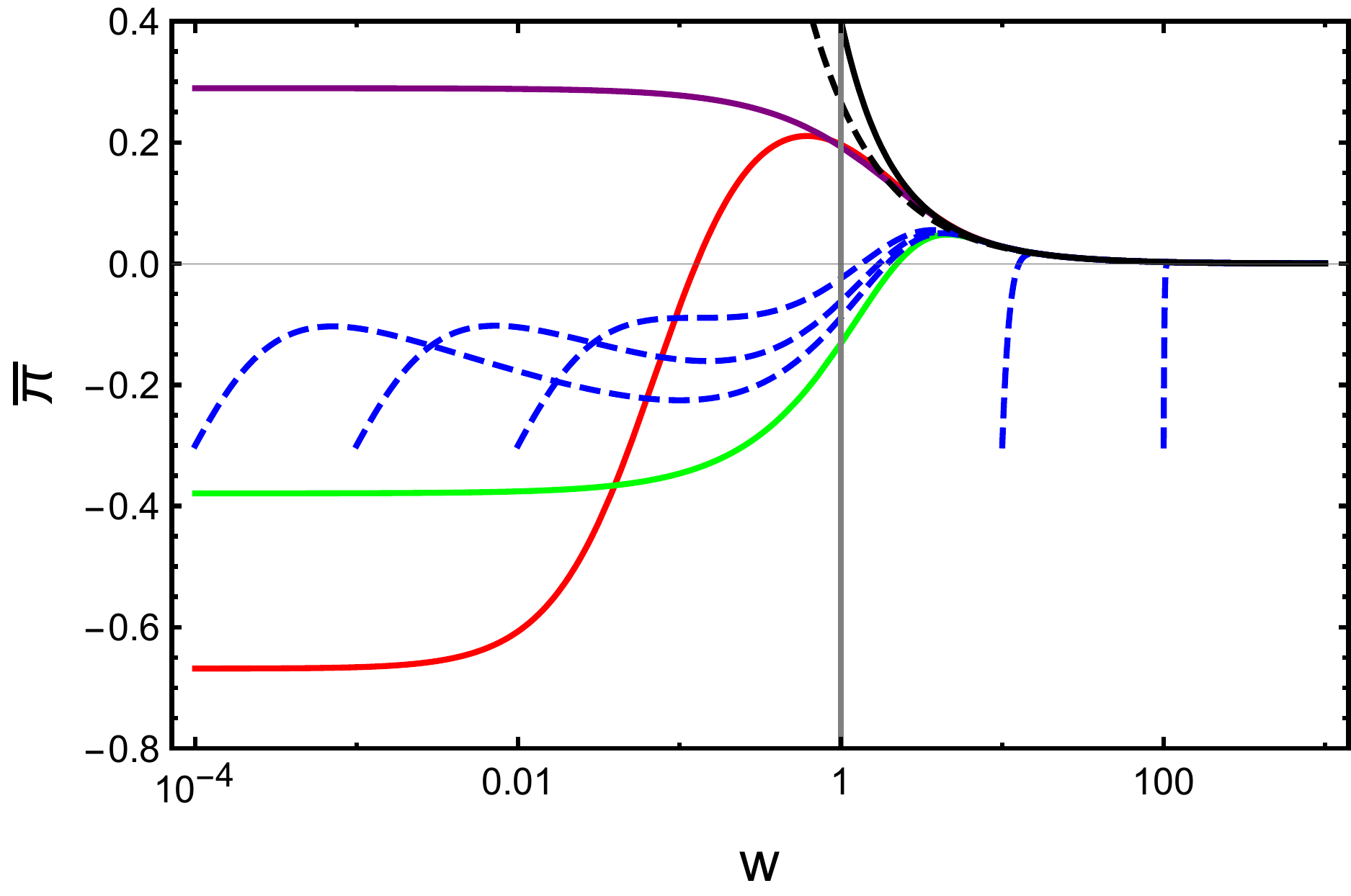}
\includegraphics[width=0.315\textwidth] {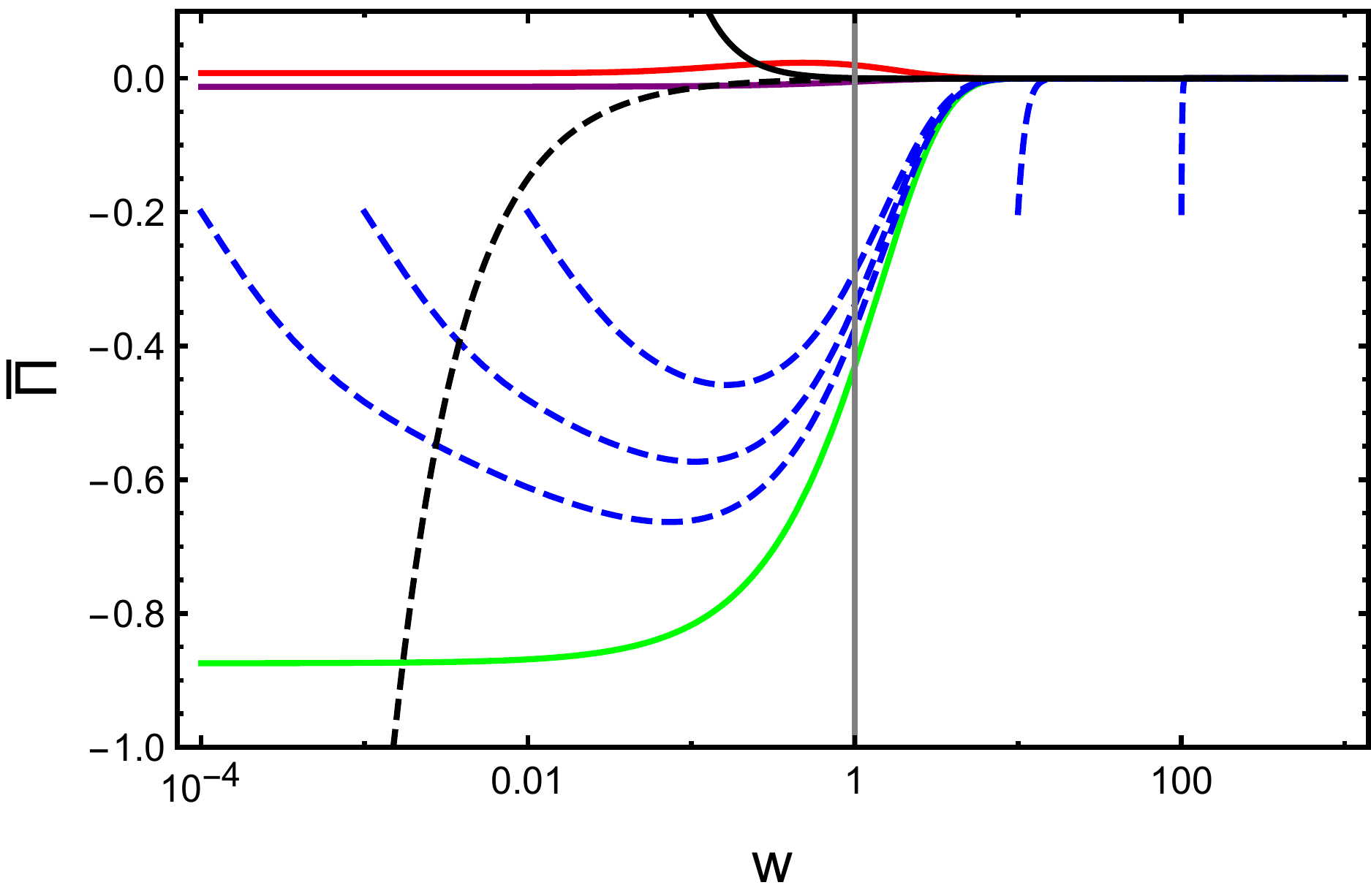}
\caption{ Numerical solutions of $g(w)$, $\bar \pi$ and $\bar \Pi$, when $\epsilon=0.01$. Blue dashed lines are solution with respect to arbitrary initial conditions starting at early and late times. Initial conditions corresponding the three free-streaming fixed points are shown as the red, purple and green solid lines. Solution from hydrodynamic gradient expansion with first order and second order gradients are shown as the black solid and dashed lines.
\label{fig:sol1}
}
\end{center}
\end{figure}

With a non-zero parameter $\epsilon$ chosen to break the conformal symmetry, we solve \Eqs{eq:eomc} numerically. Shown in \Fig{fig:sol1} are the results with respect to $\epsilon=0.01$. Solutions given arbitrary initial conditions, starting at early times with $w\ll1$ and late times with $w\gg1$, are shown as dashed blue lines. There are also solutions with respect to initial conditions corresponding to the free-streaming fixed points (red, purple and green solid lines). For comparisons, non-conformal hydrodynamic gradients to the leading order (black solid lines) and the second order (black dashed lines) are shown as well, which are known also as the solutions of NS hydro and MIS hdyro. Independent of initial conditions, at a sufficiently late time, evolutions of  $g(w)$, $\bar \pi$ and $\bar \Pi$ seem to merge to universal curves, which approach towards the expected hydrodynamic fixed points, namely, $g\to -4/3+\epsilon$ and $\bar\pi=\bar\Pi\to 0$. These universal evolution curves of $g$, $\bar\pi$ and $\bar\Pi$ at late times indicate the existence of attractor solutions in the non-conformal systems~\cite{Romatschke:2017acs,Florkowski:2017jnz,Chattopadhyay:2021ive,Jaiswal:2021uvv}. However, a more delicate analysis separating early-time and late-time evolution reveals drastically different attractor behavior in comparison with those observed in conformal fluids. Especially, we found that, while the late-time attractors are easily recognized, attractor behavior is absent at early times when $w\ll1$~\cite{Chattopadhyay:2021ive,Jaiswal:2021uvv}. This is clearly demonstrated by the fact evolutions starting at $w\ll1$ \emph{do} depend on their initial conditions in the region $w\lesssim1$.  We have also tested the numerical solutions with various values of $\epsilon$. We found that, the absence of early-time attractors is a generic feature in non-conformal fluids which prevails as long as $\epsilon>0$. 

\subsection{Gradient expansion of non-conformal hydrodynamics}
\label{sec:gexp}

Hydrodynamic gradient expansion suggests that the solution of $g(w)$, $\bar \pi$ and $\bar\Pi$ can be expanded in gradients, \ie, powers of Knudsen number,
\begin{align}
\label{eq:gexp}
g(w) \equiv \sum_{n=0}^\infty \frac{g_n}{w^n}\,,\qquad
\bar \pi(w) \equiv \sum_{n=0}^\infty \frac{k_n}{w^n}\,,\qquad
\bar \Pi(w) \equiv \sum_{n=0}^\infty \frac{q_n}{w^n}\,,
\end{align}
where $g_n$, $k_n$ and $q_n$ are constant expansion coefficients to be determined. Substituting \Eq{eq:gexp} into \Eqs{eq:eomc}, one finds equations for these coefficients,
\begin{subequations}
\label{eq:solgkq}
\begin{align}
\label{eq:solk}
&
k_{n+1} = \alpha_\pi \delta_{n0} + (n - \gamma_1) k_n + \gamma_2 q_n - \sum_{m=0}^n g_m k_{n-m} \,,\\
&
q_{n+1} = -\alpha_\Pi \delta_{n0} + \gamma_3 k_n + (n - \gamma_4) q_n - \sum_{m=0}^n g_m q_{n-m}\,,
\end{align}
\end{subequations}
where $g_n = -(4/3-\epsilon)(\delta_{n0} + q_n - k_n)$ introduces nonlinearity to the coupled equations for $k_n$ and $q_n$. 
\Eqs{eq:solgkq} can be solved iteratively order by order. For instance, by taking $n=-1$, one finds the leading order coefficients,
\be
g_0 = -4/3+\epsilon\,,\quad
k_0=q_0=0\,.
\ee
These results can be recognized in terms of the fixed points in the hydrodynamic limit of a non-conformal fluid experiencing Bjorken expansion. They also represent the expected evolution pattern of the system in the late time (close to equilibrium) limit $w\to \infty$.  The constant $g_0$, in particular, gives the relative decay rate of the the enthalpy density of an ideal non-conformal fluid, 
which approaches $-4/3+\epsilon$.  In the conformal limit $\epsilon\to0$, the decay rate reduces to the well-known $-4/3$.  Since $\bar\pi$ and $\bar \Pi$ are proportional to viscous corrections, there are no contributions from the zeroth order expansion. NS hydrodynamics is given by the first order in the gradient expansion, which can be solved by taking $n=0$ in \Eqs{eq:solgkq},
\be
k_1 = \alpha_\pi \,,\quad
q_1 = -\alpha_\Pi\,,\quad
g_1 = (4/3-\epsilon)(\alpha_\pi + \alpha_\Pi)\,. 
\ee
Note that $k_1=\alpha_\pi\propto \eta$ and $q_1=-\alpha_\Pi\propto \zeta$, as one would expect. Note also that since $\alpha_\Pi\propto \epsilon^2$, $q_1$ vanishes in the conformal limit. Actually, in the conformal limit, all the expansion coefficients of the inverse bulk Reynolds number vanish, namely, $q_n=0$. This is naturally consistent with the fact that bulk viscous pressure $\Pi=0$ in a conformal fluid.

\begin{figure}
\begin{center}
\includegraphics[width=0.315\textwidth] {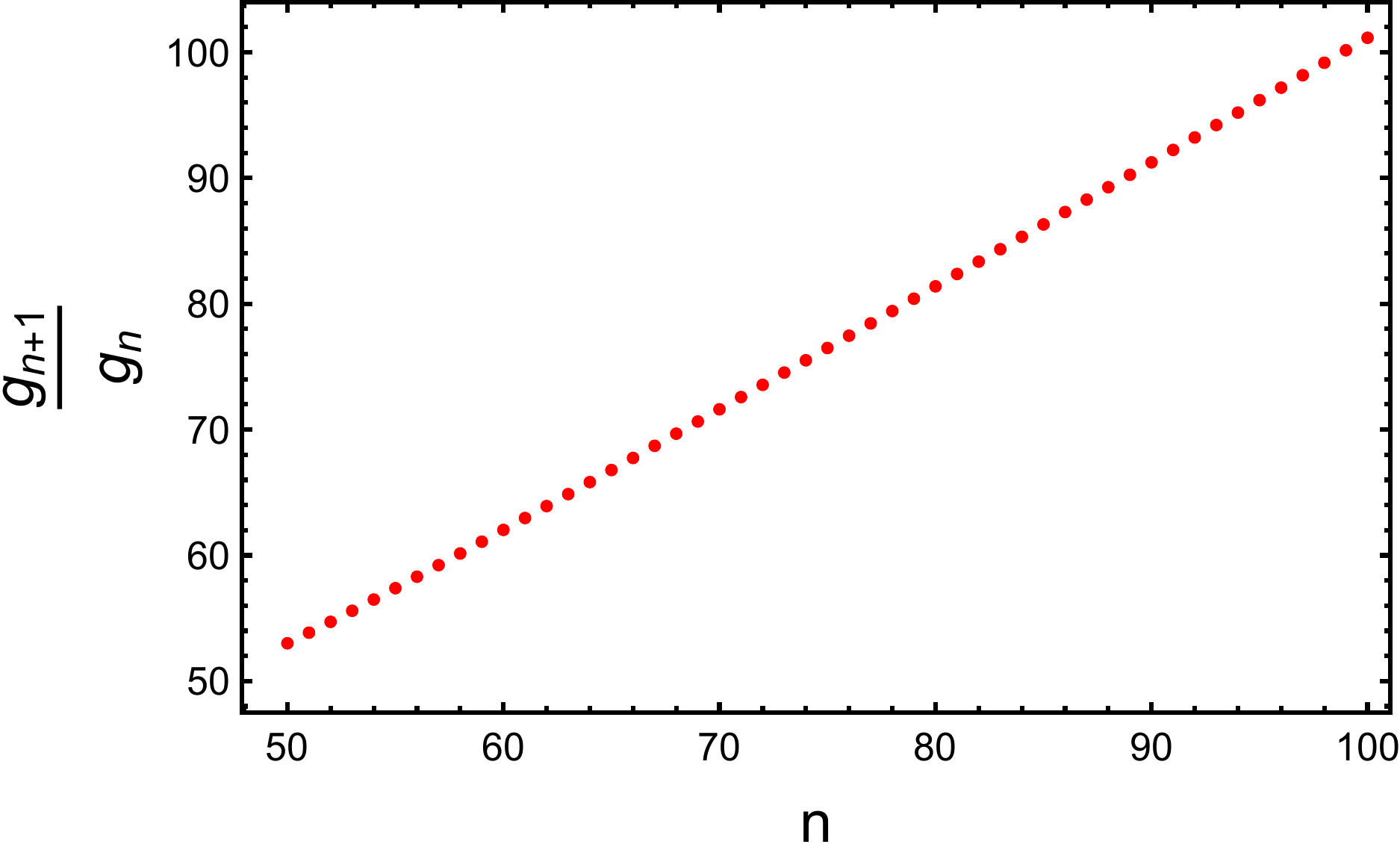}
\includegraphics[width=0.315\textwidth] {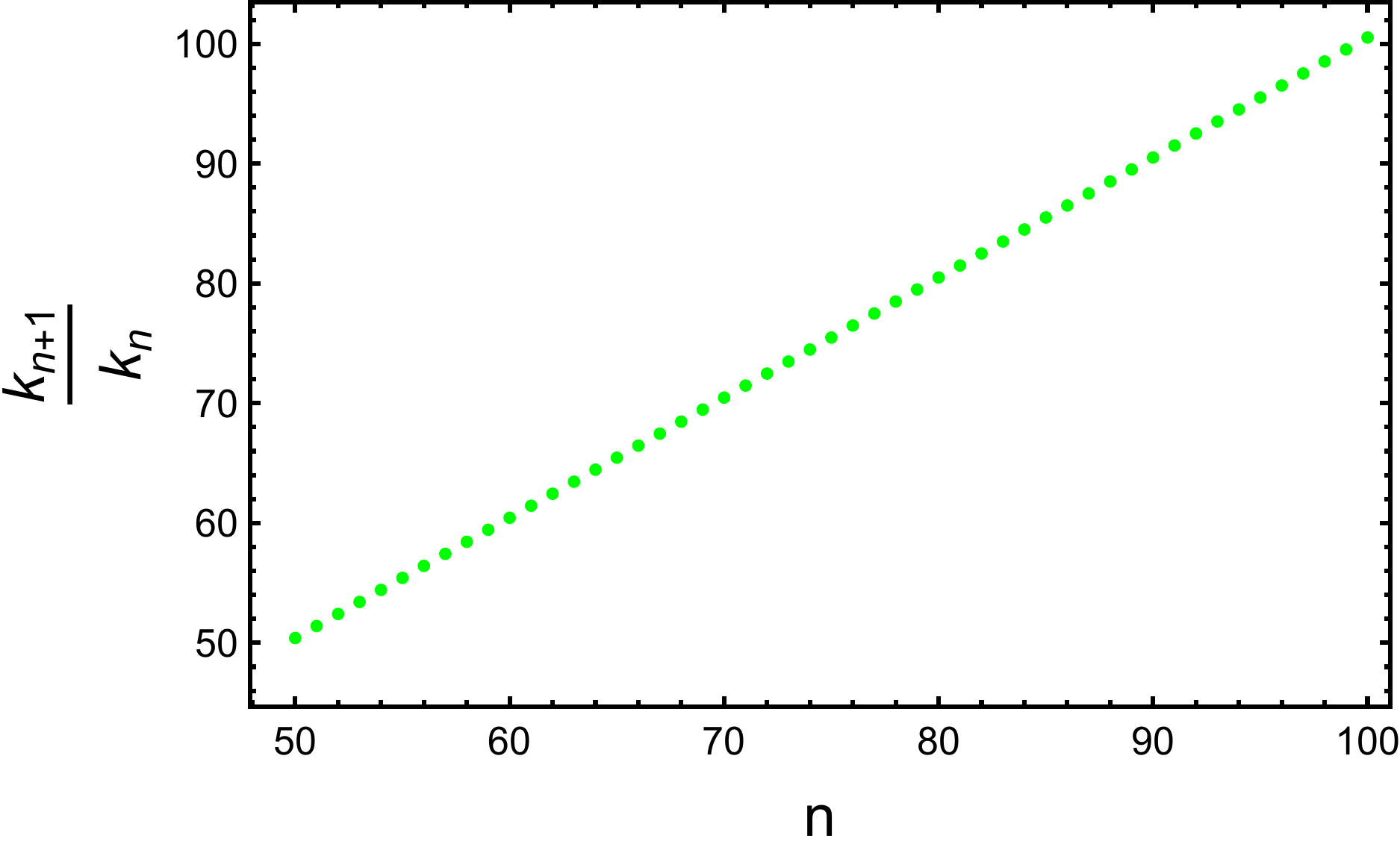}
\includegraphics[width=0.315\textwidth] {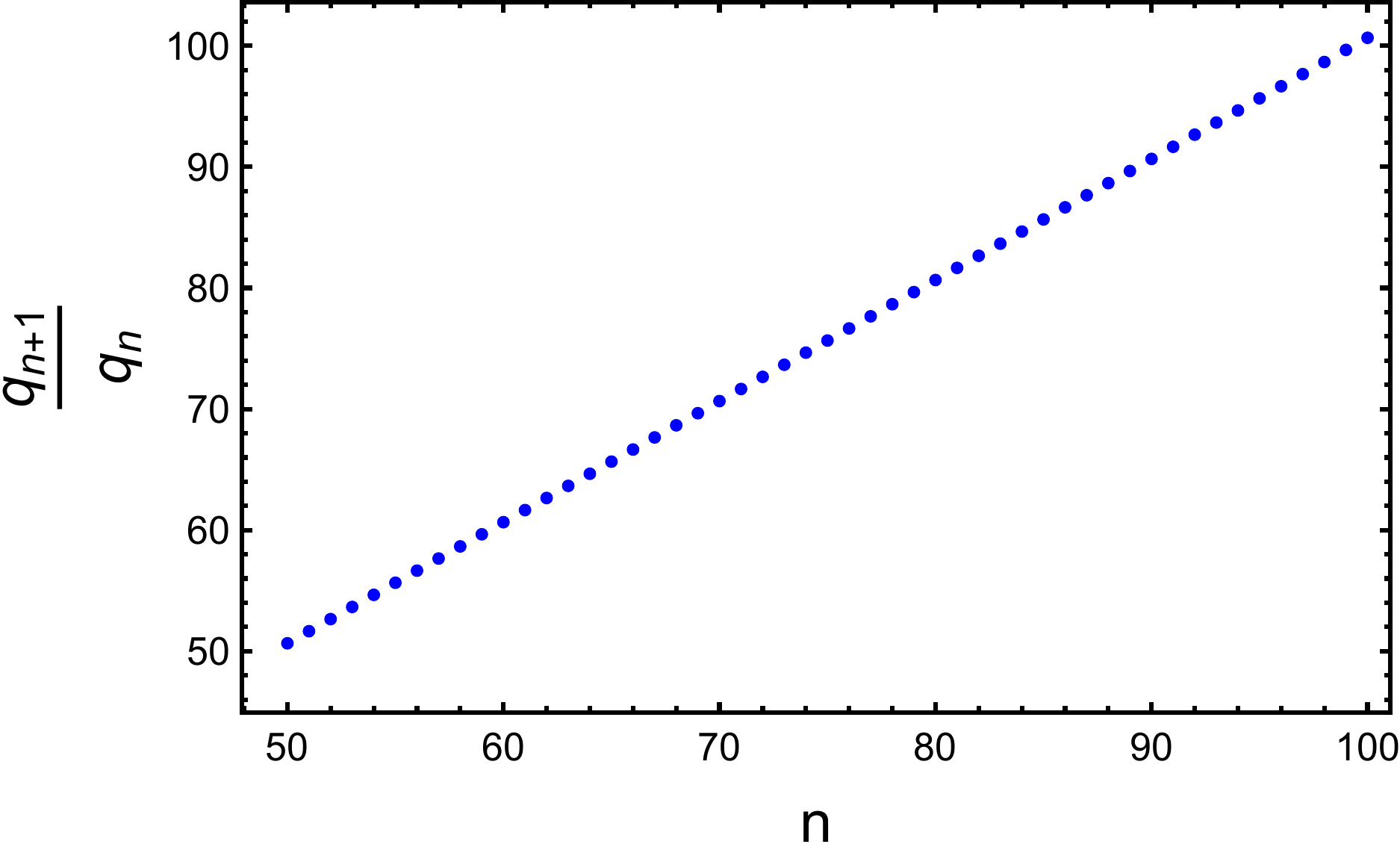}
\caption{ Ratios between coefficients of consecutive orders of the gradient expansion: $g_{n+1}/g_n$, $k_{n+1}/k_n$ and $q_{n+1}/q_n$, as a function of $n$.
\label{fig:gkq}
}
\end{center}
\end{figure}

Solutions of coefficients to arbitrary orders can be obtained similarly. In \Fig{fig:gkq}, the ratios among consecutive orders of these coefficients, $g_{n+1}/g_n$, $k_{n+1}/k_n$ and $q_{n+1}/q_n$, are shown as a function of $n$, with $\epsilon$ chosen to be 0.01. For sufficiently large $n$, these ratios are found to be linearly dependent of $n$, namely,  
when $n\to\infty$, one has
\be
\label{eq:linear}
\frac{g_{n+1}}{g_n} \sim S_g^{-1}(n+\beta_g)\,,\quad
\frac{k_{n+1}}{k_n} \sim S_k^{-1}(n+\beta_k)\,,\quad
\frac{q_{n+1}}{q_n} \sim S_q^{-1}(n+\beta_q)\,,
\ee
where $S$'s and $\beta$'s are constant parameters related to the slopes and the intercepts. 
A linear fitting in these ratios gives $S_g\approx S_k \approx S_q \approx 1$, and $\beta_g\approx \beta_k\approx \beta_q\approx 0.66$. 

Before we proceed, let us briefly recall some properties of the asymptotic gradient expansion in the conformal Bjorken flow. First, the gradient expansion of conformal fluid dynamics is asymptotic with respect to a variety of types of interactions, with the exception of hard sphere scattering~\cite{Denicol:2019lio}. Secondly, the expansion coefficients grows in a factorial fashion, which results in linearity between the ratios of expansion coefficients of consecutive orders and the expansion order. Moreover, for all the conformal fluids in which the gradient expansion is asymptotic, the slope $S>0$ and the intercept parameter $\beta<0$. It has also been noticed that, when relaxation time $\tau_\pi$ is parameterized in terms of proper time through a constant $\Delta$, \ie, $\tau_\pi \sim \tau^{1-\Delta}$ as in \Eq{eq:relax}, the slope can actually be identified as $S=\Delta^{-1}$~\cite{Blaizot:2021cdv}. Thirdly, for conformal fluids, Borel resum technique can be applied to the asymptotic series, leading  to a tran-series solution. Attractor solution corresponds to resummation of the tran-series with respect to one particular initial condition~\cite{Blaizot:2020gql}. All solutions other than the attractor contain transient modes, whose evolution is controlled by an extra factor $e^{-Sw}w^\beta$.  Sicne $S>0$ and $\beta<0$, all the transient solutions eventually decay towards the attractor. In particular, at early times $w\ll1$, it is dominantly a power-law decay $w^\beta$, while at late times $w\gg1$ it is an exponential decay $e^{-Sw}$, corresponding to the early-time and late-time attractor behavior respectively. 

For the non-conformal fluids, we thereby understand that, the linear growth of the ratios among coefficients of consecutive orders shows equivalently a factorial growth of the expansion coefficients, which further signifies that the gradient expansions of $g(w)$, $\bar\pi$ and $\bar\Pi$ are all asymptotic. These asymptotic series are naturally generalizations of what has been discovered in conformal fluids. Because $\Delta_\pi=\Delta_\Pi=1$ has been taken into account in the numerical calculations, the slopes in \Eq{eq:linear} are indeed unity, as expected. However, unlike conformal fluids, the intercept parameters $\beta$'s \emph{can be} positive. 


With respect to \Eqs{eq:solgkq}, the linear growth of the ratios can also be verified analytically. For instance, divide both sides in \Eq{eq:solgkq} by $k_n$ and $q_n$ respectively, and consider in the limit $n\to \infty$, one finds~\cite{Basar:2015ava},
\begin{subequations}
\label{eq:linear2}
\begin{align}
\label{eq:linear2a}
\frac{k_{n+1}}{k_n} \sim& n - \gamma_1 + \gamma_2 \frac{q_n}{k_n} - g_0 + O\left(\frac{1}{n}\right)\,,\\
\label{eq:linear2b}
\frac{q_{n+1}}{q_n} \sim& n - \gamma_3 \frac{k_n}{q_n} + \gamma_4 - g_0 + O\left(\frac{1}{n}\right)\,,
\end{align}
\end{subequations}
which indicates exactly $S_k=S_q=1$. Since \Eqs{eq:linear2} are obtained independent of $\epsilon$, one realizes that
whether the linear relations in \Eqs{eq:linear} 
emerge, and also the fact that slopes are unity, do not depend on the symmetry broken paramter $\epsilon$. 
The independence on $\epsilon$ can be also tested numerically, by repeating the above analysis with various values of $\epsilon$. As a consequence, one may conclude that the gradient expansion in a non-conformal boost-invariant plasma being asymptotic is a feature irrespective of the broken conformal symmetry. When $\epsilon\to0$, accompanied by the expectation that $q_n\to0$, \Eqs{eq:linear} reduce smoothly to the linear relation in a conformal fluid~\cite{Basar:2015ava}. In this case, one finds $\beta_k=-\gamma_1-g_0=-10/21<0$. When conformal symmetry is broken via $\epsilon>0$, it becomes more involved to extract $\beta$'s from \Eqs{eq:linear2}, although in principle one would expect a smooth transition of $\beta_k$ from negative to positive values. In particular, in some certain region of small but nonzero $\epsilon$, $\beta_k$ is still expected negative.

\Eqs{eq:linear2} can be simply generalized to arbitrary values of $\Delta_\pi$ and $\Delta_\Pi$, corresponding to more sophisticated parameterizations of the relaxation times. In such scenarios, the results  differ only in the restoration of $\Delta_\pi$ and $\Delta_\Pi$ in front of $n$ in \Eq{eq:linear2a} and \Eq{eq:linear2b}, respectively. Except in the case of hard sphere scatterings, where $\Delta_\pi=\Delta_\Pi=0$, the linearities in the ratios are not affected, neither does the asymptotic nature of the non-conformal gradient expansion.

It would also be interesting to carry out Borel sum with respect to the asymptotic series in \Eq{eq:gexp}, in analogous to the analysis for conformal fluids. 
Following the standard procedure of Borel sum, one expects the slope parameters $S$'s to correspond to the leading singularities in the complex Borel plane, which eventually gives rise to the exponential factor $e^{-Sw}$. On the other hand, the intercept parameter $\beta$'s should appear in the power law factor $w^\beta$, after an inverse Laplace transform with respect to the Borel transform of the original asymptotic series. However, as we have noticed, $\beta_k$ could be negative for finite $\epsilon$, which apparently conflicts to the observation that early-time attractor is absent in the non-conformal fluids as long as $\epsilon>0$. This, as to be explained in the next section, is due to the fact that the two independent variables, $\bar\pi$ and $\bar\Pi$, are actually mixtures of stable and instable modes. 

\subsection{Early-time stable and instable modes}
\label{sec:modes}

To examine the coupling of $\bar\pi$ and $\bar\Pi$, and to investigate the evolution of  initial condition dependent modes, we perturb the solution around hydrodynamic gradient expansion. To do so, we first rewrite \Eqs{eq:eomc} as follows,
\begin{subequations}
\label{eq:eomN}
\begin{align}
	\frac{\partial\t{\pi}}{\partial\ln w}&=\sqrt{\gamma_3}\alpha_\pi-\gamma_1\t{\pi}+\sqrt{\gamma_2\gamma_3}\t{\Pi}+(1+c_s^2)\left(1+\frac{\t{\Pi}}{\sqrt{\gamma_3}}-\frac{\t{\pi}}{\sqrt{\gamma_2}}\right)\t{\pi}-w\t{\pi}\,,\\
		\frac{\partial\t{\Pi}}{\partial\ln w}&=-\sqrt{\gamma_2}\alpha_\Pi+\sqrt{\gamma_2\gamma_3}\t{\pi}-\gamma_4\t{\Pi}+(1+c_s^2)\left(1+\frac{\t{\Pi}}{\sqrt{\gamma_3}}-\frac{\t{\pi}}{\sqrt{\gamma_2}}\right)\t{\Pi}-w\t{\Pi}\,,
\end{align}
\end{subequations}
where for later convenience, we have redefined $\bar\pi\to \t \pi = \sqrt{\gamma_3} \bar \pi$ and $\bar\Pi\to\t \Pi = \sqrt{\gamma_2} \bar \Pi$ to symmetrize the equations. 

\begin{figure}[h]
\begin{center}
\includegraphics[width=0.55\textwidth] {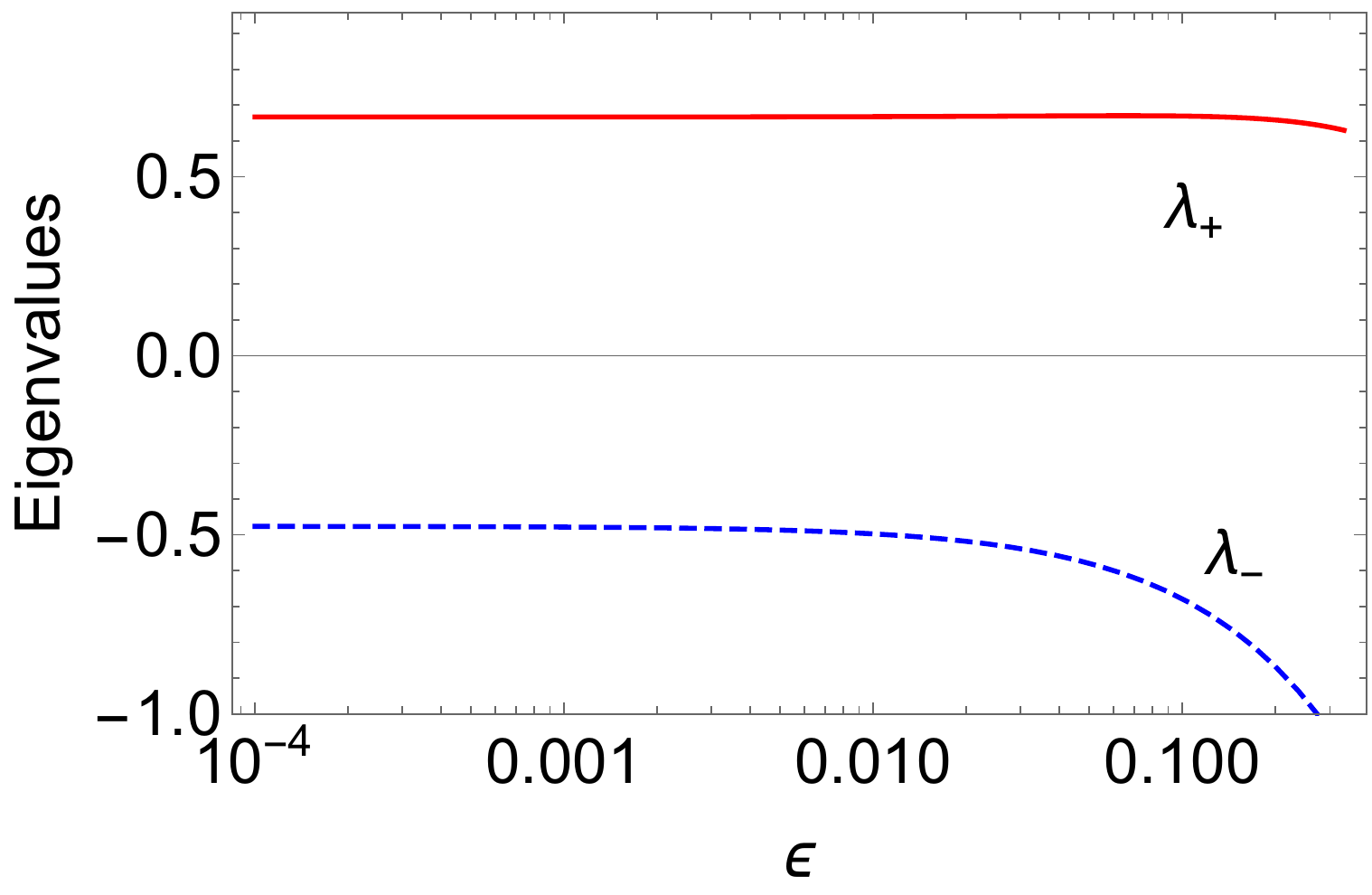}
\caption{ Eigenvalues of the free-streaming matrix $\Lambda$ as a function of $\epsilon=1/3-c_s^2$. When $\epsilon\to 0$, the systems of $\t\pi$ becomes decoupled and the eigenvalue approaches its conformal correspondence, $\lambda_-\to -10/21$.
\label{fig:lambdas}
}
\end{center}
\end{figure}

\Eqs{eq:eomN} can be treated as 
nonlinear ordinary differential equations for the vector $\qket\psi=(\t \pi, \t \Pi)=({\t \pi}_{\rm hydro} + \delta \t \pi, {\t \Pi}_{\rm hydro} + \delta \t \Pi)$. Here, $\t \pi_{\rm hydro}$ and $\t\Pi_{\rm hydro}$ correspond to the solution given by hydrodynamic gradient expansion. It should be noted that, only when $\epsilon=0$, namely, in the conformal limit, do $\t \pi$ and $\t \Pi$ in \Eqs{eq:eomN} become decoupled. Since $\t \pi_{\rm hydro} \sim \t \Pi_{\rm hydro} \sim 1/w$, to the lowest order in gradients and linearized order of perturbations, one finds the equations for $(\delta\t\pi,\delta\t\Pi)$,
\begin{subequations}
\begin{align}
\frac{d \delta \t \pi}{d\ln w} &= (1+c_s^2-\gamma_1) \delta \t\pi + \sqrt{\gamma_2\gamma_3} \t \delta \Pi - w \delta \t \pi\,,\\
\frac{d \delta \t \Pi}{d\ln w} & = \sqrt{\gamma_2\gamma_3}\delta \t \pi +(1+c_s^2-\gamma_4) - w \delta \t \Pi\,,
\end{align}
\end{subequations}
or formally,
\be
\label{eq:eompsi}
\frac{d }{d \ln w}\qket{\delta \psi} = \Lambda \qket{\delta \psi} - w \qket{\delta \psi}
\ee
with
\be
\qket{\delta \psi} =
\begin{pmatrix}
\delta \t \pi\\
\delta \t \Pi
\end{pmatrix}
\qquad\mbox{and}\qquad
\Lambda=
\begin{pmatrix}
4/3-\epsilon - \gamma_1 & \sqrt{\gamma_2\gamma_3}\\
\sqrt{\gamma_2\gamma_3} & 4/3-\epsilon -\gamma_4
\end{pmatrix}
\ee
Here the matrix $\Lambda$ plays a similar role as $H_0$ in \Eq{eq:matrix}, describing the free-streaming dynamics at early times. 
In the standard basis, $\qket{e_1}=(1,0)$ and $\qket{e_2}=(0,1)$,
one has $\delta\t\pi = \qbra{e_1}\delta \psi\ket$ and $\delta \t\Pi = \qbra{e_2} \delta\psi\ket$.
The symmetric matrix $\Lambda$ allows for orthogonal eigenvectors, satisfying $\Lambda \qket{\phi_\pm} = \lambda_\pm \qket{\phi_\pm}$ and $\qbra{\phi_+}\phi_-\ket=0$. The eigenvalues and the eigenvectors are functions of the symmetry broken parameter $\epsilon$ and the dimensionless second order transport coefficients, $\gamma_1$, $\gamma_2$, etc.. For instance, eigenvalues are
\be
\lambda_\pm(\epsilon) =
\frac{1}{6}\left[8-6\epsilon-3(\gamma_1+\gamma_4)\pm3\sqrt{(\gamma_1-\gamma_4)^2-2 \gamma_2\gamma_3}\right]\,,
\ee
whose dependence on $\epsilon$ is shown in \Fig{fig:lambdas}.  Expanding in these eigenvectors, one has
\be
\qket{\delta\psi(w)}= \xi_+(w)\qket{ \phi_+} + \xi_-(w)\qket{ \phi_-}\,,
\ee
so that \Eq{eq:eompsi} reduces to equations for the decoupled eigen-modes,
\be
\frac{d \xi_\pm(w)}{d\ln w} = \lambda_\pm \xi_\pm - w\xi_\pm\,,
\ee
which can be solved as $
\xi_\pm(w) \sim w^{\lambda_\pm} e^{-w}$.
In both modes, the same exponential suppression factor contributes and becomes dominant when $w\gg1$, so that the late-time attractor behavior is guaranteed. As shown in \Fig{fig:lambdas}, with respect to the increase of $\epsilon$, the eigenvalues always satisfy $\lambda_+>0$ and $\lambda_-<0$. The negative eigenvalue corresponds to a stable mode, which leads to a power-law decay of initial condition dependent perturbations at early times and thus the early-time attractor behavior. However, the mode with positive eigenvalue $\lambda_+$ results in early-time instability, which destroys the early-time attractor behavior. As a result, as mixtures of the stable and instable modes, for the solution of perturbations in both the shear and the bulk channels,
\begin{subequations}
\label{eq:xitopi}
\begin{align}
\delta\t\pi(w) = \xi_+(w)\qbra{e_1}\phi_+\ket+\xi_-(w)\qbra{e_1}\phi_-\ket\,,\\
\delta \t\Pi(w) = \xi_+(w)\qbra{e_2}\phi_+\ket+\xi_-(w)\qbra{e_2}\phi_-\ket\,,
\end{align}
\end{subequations}
there are no early-time attractors present, even though these perturbations do vanish at late time which leads to late-time attractor.

\section{Restoration of non-conformal early-time attractor}
\label{sec:4}

In previous sections, we have verified that the absence of early-time non-conformal attractor is a generic feature in the non-conformal boost invariant fluids, as essentially a consequence of the instable mode which grows according to a power-law factor $w^{\lambda_+}$. With respect to realistic simulations of heavy-ion collisions, the absense of early-time attractorr would forbid early-time  initialization of hydrodynamics when $w\lesssim 1$. In terms of proper time, this condition implies that hydrodynamic modeling of the QGP cannot be valid before $\tau\sim\tau_R$. To resolve the discrepancy, one needs to restore the early-time attractor behavior in non-conformal fluids. Here we propose two possible strategies. 

\subsection{Mixtures among the shear and bulk channels}

\begin{figure}
\begin{center}
\includegraphics[width=0.46\textwidth] {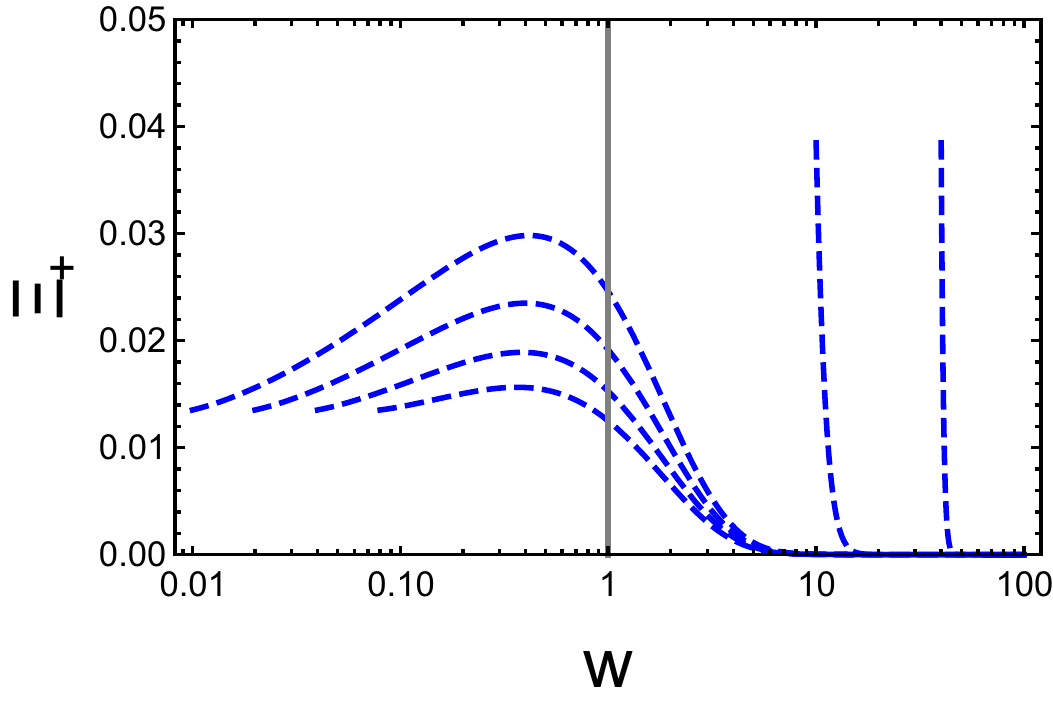}
\includegraphics[width=0.47\textwidth] {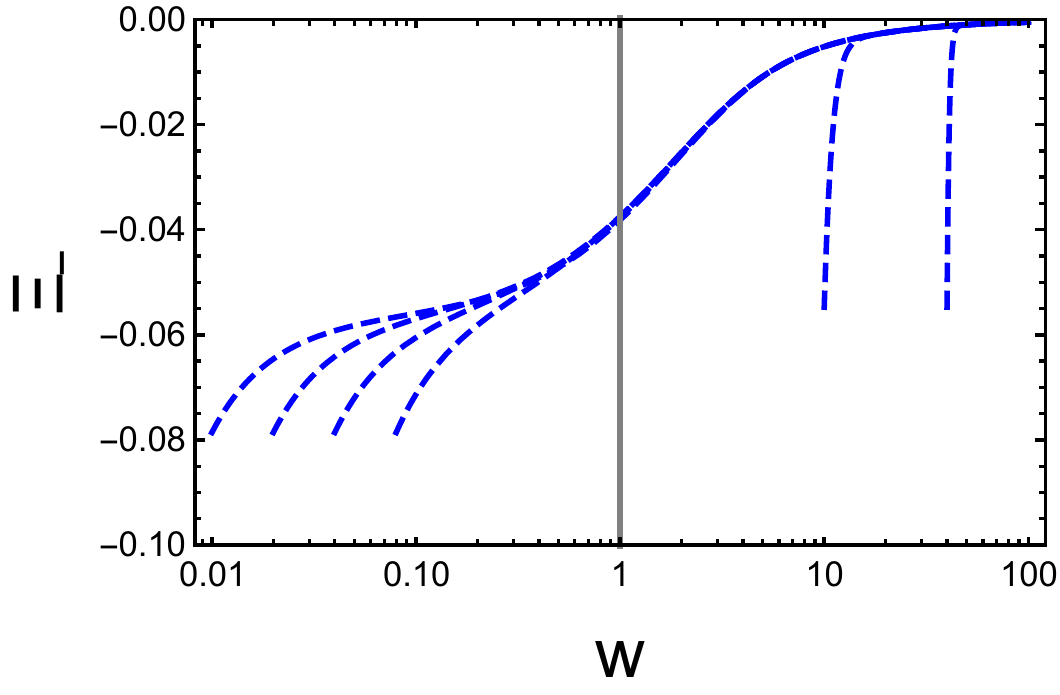}
\caption{Numerical solutions of $\Xi_\pm$ with respect to various initial conditions, starting at early times with $w\ll1$ and late times with $w\gg1$, as functions of $w$. 
\label{fig:Xis}
}
\end{center}
\end{figure}

Upon a reverse transformation, \Eqs{eq:xitopi} implies the existence of mixtures among the shear and the bulk channels,
\begin{subequations}
\label{eq:pitoxi}
\begin{align}
\Xi_+(w) = \t\pi(w)\qbra{e_1}\phi_+\ket+\t\Pi(w)\qbra{e_2}\phi_+\ket \,,\\
\Xi_-(w) = \t\pi(w)\qbra{e_1}\phi_-\ket+\t\Pi(w)\qbra{e_2}\phi_-\ket \,,
\end{align}
\end{subequations}
which defines two new dimensionless quantities $\Xi_\pm(w)$. Note that these are mixtures of the full shear and bulk inverse Reynolds numbers, not perturbations. Physically, evolution of these quantities mixes evolution of the shear and the bulk viscous corrections, with the mixing determined by constants $\qbra{e_1}\phi_+\ket$, $\qbra{e_2}\phi_+\ket$, etc. Note that these mixing constants are the components of the eigenvectors, as function of second order transport coefficients and symmetry breaking parameter $\epsilon$. To the linearized order of perturbations around attractor solutions, by construction, $\Xi_+(w)$ is expected early-time instable (without early-time attractor) while $\Xi_-$ is expected early-time stable (with early-time attractor). The restoration of early-time attractor in $\Xi_-$ can be seen also as the detailed cancellation among the instable modes in the shear and the bulk channels. On the contrary, $\Xi_+$ is purely instable at early times where stable modes are entirely canceled. The late-time behavior in $\Xi_\pm$ is expected similar to that of the shear and the bulk channels.

Numerical solutions of these two quantities are shown in \Fig{fig:Xis}, with respect to some arbitrary initial conditions starting at $w\ll1$ and $w\gg1$.  Although the late-time attractor is seen in both $\Xi_+$ and $\Xi_-$, as being demonstrated as the universal evolution pattern $\Xi_\pm\to 0$ when $w\gg1$, early-time attractor behavior exhibits only in $\Xi_-$\footnote{
In \Ref{Chattopadhyay:2021ive},  on a different ground, early-time non-conformal attractor is realized in the mixture of the shear and bulk viscous corrections in the form $-\pi/\P+\Pi/\P$, while for the case of $\epsilon=0.01$, for comparison we have $\Xi_-\approx -0.18\pi/(\E+\P)+0.13\Pi/(\E+\P)$.
}.  

It should be emphasized that the restoration of the early-time attractor solution in $\Xi_-$ is only valid to the linearized order of initial perturbations. Therefore, the relation in \Eq{eq:pitoxi} is only an approximation for cases when $w$ is not too small, otherwise contributions from higher orders, e.g., $(w^{\lambda_+})^2$, can be significant. 

\subsection{Early-time non-conformal attractor with revised $\gamma_4$}

\begin{figure}
\begin{center}
\includegraphics[width=0.315\textwidth] {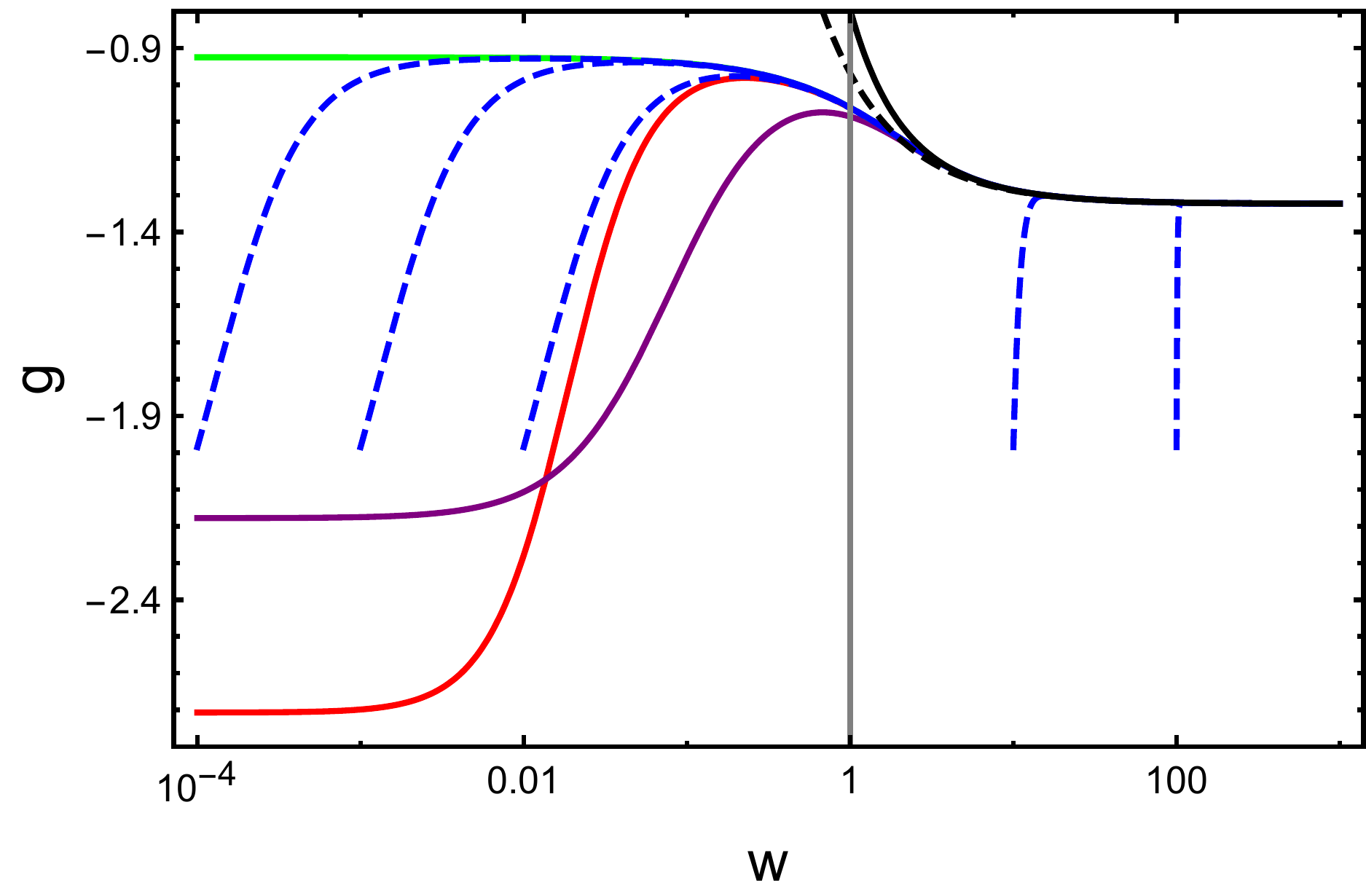}
\includegraphics[width=0.315\textwidth] {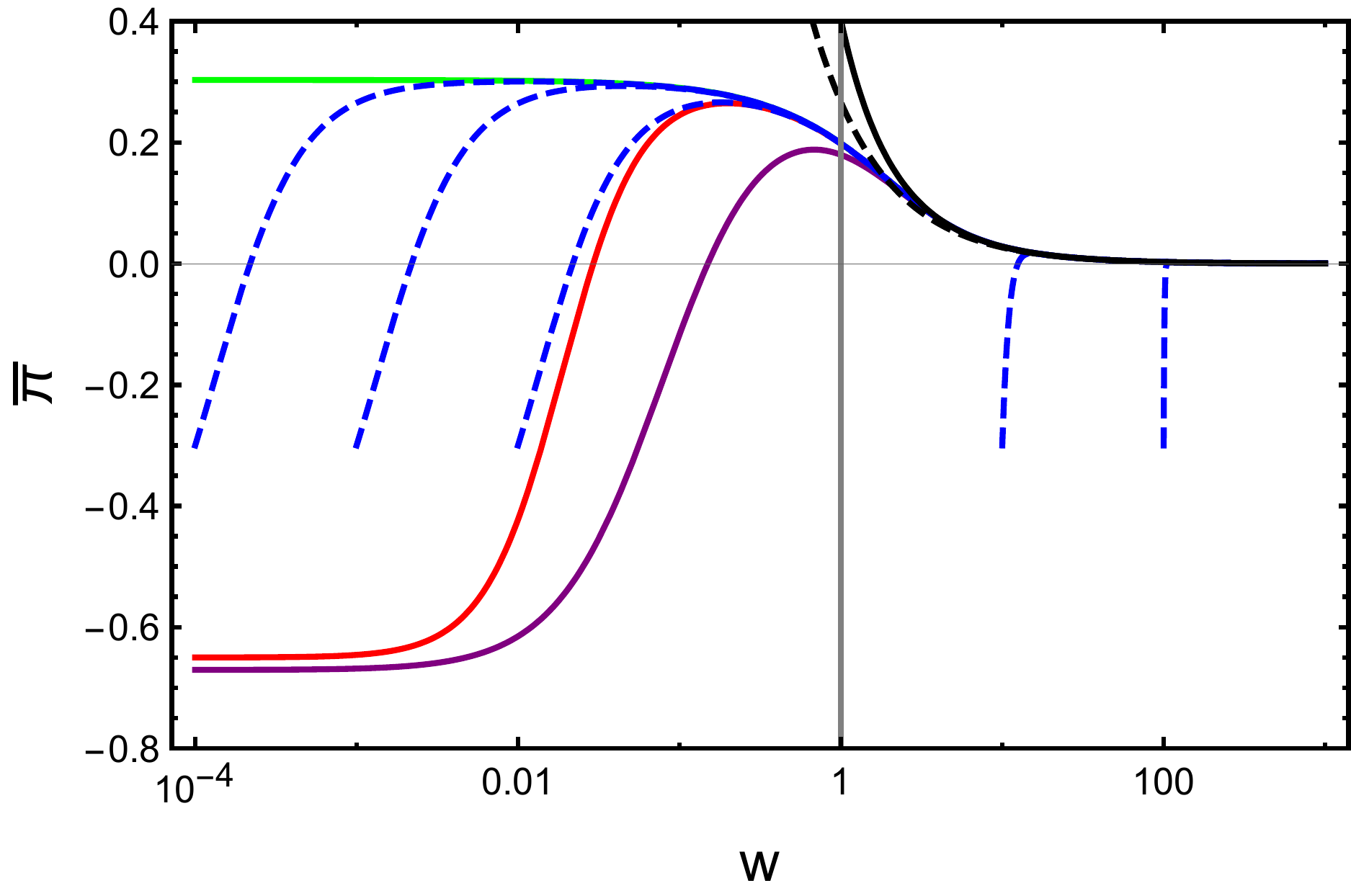}
\includegraphics[width=0.315\textwidth] {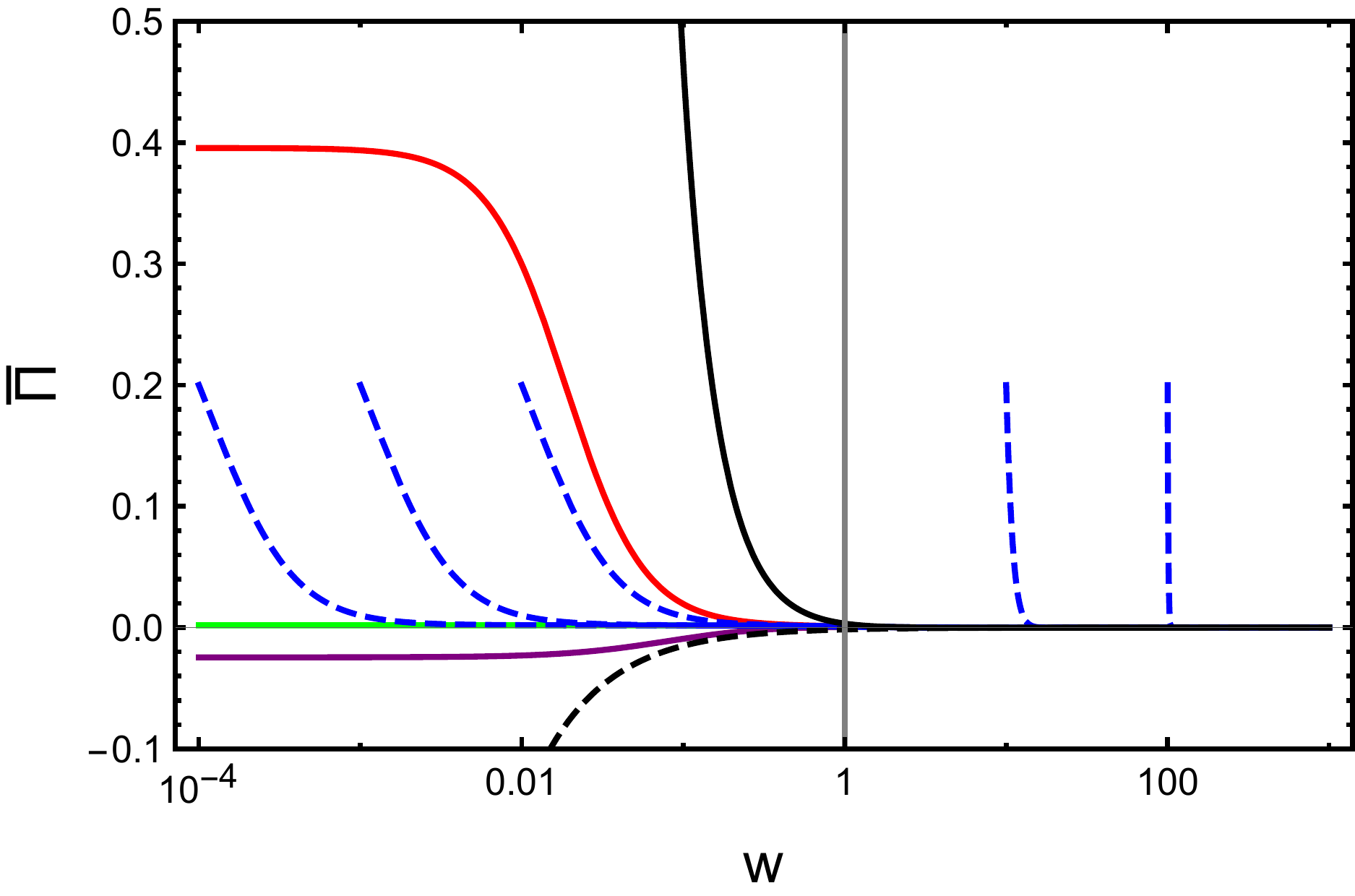}
\caption{  Numerical solutions of $g(w)$, $\bar \pi$ and $\bar \Pi$ with a revised $\gamma_4=3/2$, when $\epsilon=0.01$. Blue dashed lines are solution with respect to arbitrary initial conditions starting at early and late times. Initial conditions corresponding the three free-streaming fixed points are shown as the red, purple and green solid lines. Solution from hydrodynamic gradient expansion with first order and second order gradients are shown as the black solid and dashed lines.
\label{fig:sol2}
}
\end{center}
\end{figure}


The aforementioned instable mode is the key factor that destroys the early-time non-conformal attractor. This instable mode is due to a positive eigenvalue in the free-streaming matrix $\Lambda$.
If both eigenvalues of the free-streaming matrix are made non-positive, the resulted off-equilibrium evolution and the non-conformal attractor would be well established, either at early times or at late times. 

Mathematically, accounting for the parameter dependence in the free-streaming matrix $\Lambda$, non-positive eigenvalues can be achieved in certain regions of the parameter space of $\gamma_1$, $\gamma_2$, $\gamma_3$ and $\gamma_4$, corresponding to some particular values of the second order transport coefficients. However, whether these desired values of transport coefficients are physically allowed needs to be investigated. 
The constraint comes directly from the consideration that there should be a smooth transition to the conformal fluids, once the symmetry broken parameter reduces to zero. By examining the matrix structure of $\Lambda$, one first notices that as $\epsilon\to0$, the off-diagonal components vanish continuously. This is actually understandable on a physics ground, regarding the fact the off-diagonal components depend on parameters $\gamma_2$ and $\gamma_3$, which are further related to the effects of shear-bulk coupling in hydrodynamics.  In the conformal limit, the matrix becomes diagonalized, with its eigenvalues approach
\be
\label{eq:conEn}
\lambda_-\to \frac{4}{3} - \gamma_1 = -\frac{10}{21}\,,\qquad
\lambda_+\to \frac{4}{3} - \gamma_4 = \frac{2}{3}\,.
\ee
Since $\lambda_-\to-10/21$ is the exact conformal correspondence of the eigenvalue which characterizes the early-time conformal attractor, $\gamma_1$ cannot be modified.  For a sufficiently small but nonzero $\epsilon$, it can be shown that corrections to the eigenvalues in \Eq{eq:conEn} are of the order of $\epsilon$, arising from the off-diagonal components of the matrix. This small correction does not affect the sign of $\lambda_\pm$, thus $\gamma_2$ and $\gamma_3$ are not relevant in determining the eigenvalues. Therefore, it is only $\gamma_4$ that should be revised to recover the non-conformal early-time attractor. 


Given these simple arguments, one can deduce the following equation that determines the critical value of $\gamma_4$ as a function of $\epsilon$,
\be
\lambda_+(\epsilon,\gamma_4^{\rm cri}) = 0\,,
\ee
which can be solved as
\be
\gamma_4^{\rm cri} = \frac{\delta_{\Pi\Pi}^{\rm cri}}{\tau_\Pi} = \frac{1000+342 \epsilon-1575 \epsilon^2}{75(10+21 \epsilon)}
=\frac{4}{3} -\frac{293}{125}\epsilon+\frac{1764}{625}\epsilon^2 + O(\epsilon^3)\,.
\ee
Note that in the pure conformal limit, the critical value of $\gamma_4$ (or $\delta_{\Pi\Pi}$) is necessisarily twice of the original one obtained from kinetic theory. 

By taking the revised value of $\gamma_4$, and repeating the analyses of the pre-equilibrium evolution of the non-conformal fluid, one finds solutions of $g(w)$, $\bar\pi$ and $\bar\Pi$, as shown in \Fig{fig:sol2}. Here again, $\epsilon=0.01$ is taken for the purpose of illustration. With a $\gamma_4$ taken greater than the corresponding critical value, solutions with both early-time and late-time attractors are indeed recovered. It is also interesting to note that, the attractor solutions in $g(w)$, $\bar\pi$ and $\bar\Pi$ can be well recognized as the solution starting the free-streaming stable fixed point (green lines in \Fig{fig:sol2}). This is similar to the attractor in conformal fluids, which connects the free-streaming fixed point and the hydrodynamic fixed point.

\section{Summary and discussion}    
\label{sec:5}

We have applied the non-conformal second order viscous hydrodynamics to a boost-invariant plasma out of equilbrium. With respect to various initial conditions, out-of-equilibrium system evolution can be fully characterized by the relative decay rate of enthalpy density, the inverse shear and the bulk Reynolds numbers. From the numerical solutions of these quantities, out-of-equilibrium non-conformal systems is found to evolve towards an ideal fluid at late times, following some universal patterns. These universal evolutions indicate the existence of non-conformal hydrodynamic attractor. However, in comparison to the conformal attractor solution, it is drastically different since there is not early-time attrator in non-conformal fluids. 

The non-conformal attactor solutions can be investigated in many different aspects. Especially, as has been shown in the current study, the solutions to non-conformal hydrodynamics can be expanded in gradients. Independently, these are gradient expansions of the inverse shear and bulk Reynolds numbers, both of which are asymptotic. 
In analogous to the analysis of conformal fluids, applying Borel summation to these non-conformal asymptotic series in the shear and the bulks channel results in an exponential factor that dominates the late-time evolution of transient modes, which explains the observed late-time attractor behavior in numerical solutions. However, at early times, 
the power-law decay factor that has been extracted 
in conformal fluids cannot be identified in the non-conformal fluid. In fact, by perturbing around the solution of gradient expansion, the coupled variables $\bar\pi$ and $\bar\Pi$ are found as mixtures of an early-time stable mode and an early-time instable mode. Especially, the early-time instable mode, which is rooted in a positive eigenvalue of the free-streaming matrix, is the key factor that destroys the early-time non-conformal attractor.

The early-time attractor behavior is necessary to practical simulations of hydrodynamics in heavy-ion collisions, which allows for initialiation of hydrodynamics at very early stages of QGP, especially, for the QGP medium generated in small colliding systems. To restore the early-time non-conformal attractor for realistic QGP, there are two possible strategies. It can be realized in one specified form of admixture among the shear and the bulk channels, where the early-time instable modes cancel approximately. The early-time non-conformal attractor can be also obtained when the quadratic coupling between the bulk viscous pressure and the expansion rate is enlarged through the revised transport coefficient $\delta_{\Pi\Pi}$. 

In this current study, for simplicity we have chosen the same constant relaxation time to capture interactions in the shear and the bulk channels, through the parameterization parameters $\Delta_\pi=\Delta_\Pi=1$. However, as has been discussed previously, more involved interactions using different parameterizations of the relaxation times do not change the qualitative features of the non-conformal attractor. Especially, the absence of early-time non-conformal attractor is due to involvement of early-time instable mode, which is further originated from one positive-valued eigen-mode from the free-streaming dynamics. Therefore, the early-time attractor behavior in the non-conformal fluids does not depend on interactions.

\acknowledgements

L.Y. is grateful for fruitful discussions with Jean-Paul Blaizot and Yi Yin. 
This work is supported by National Natural Science Foundation of China under contract number 11975079.

\bibliography{refsbib}
\end{document}